\documentclass[12pt]{article}
\usepackage{a4wide}

\usepackage{graphics}

\newcommand{\TeV}{\mbox{ TeV}}
\newcommand{\GeV}{\mbox{ GeV}}
\newcommand{\stopp}{\ensuremath{\tilde t}}
\newcommand{\sbottom}{\ensuremath{\tilde b}}
\newcommand{\hplus}{\ensuremath{H^+}}

\newcommand{\mHp}{\ensuremath{M_{H^\pm}}}
\newcommand{\mt}{\ensuremath{m_t}}
\newcommand{\mh}{\ensuremath{M_{h^0}}}
\newcommand{\mH}{\ensuremath{M_{H^0}}}
\newcommand{\mA}{\ensuremath{M_{A^0}}}
\newcommand{\mw}{\ensuremath{M_W}}

\newcommand{\afb}{\ensuremath{A_{FB}}}
\newcommand{\tb}{\ensuremath{\tan\beta}}
\newcommand{\ssw}{\ensuremath{\sin ^2\theta_W}}

\newcommand{\lsim}{\mbox{ \raisebox{-4pt}{${\stackrel{\textstyle <}{\sim}}$} }}

\newcommand{\figtree}{
\begin{figure}[tbp]
\begin{center}
\resizebox{10cm}{!}{\includegraphics{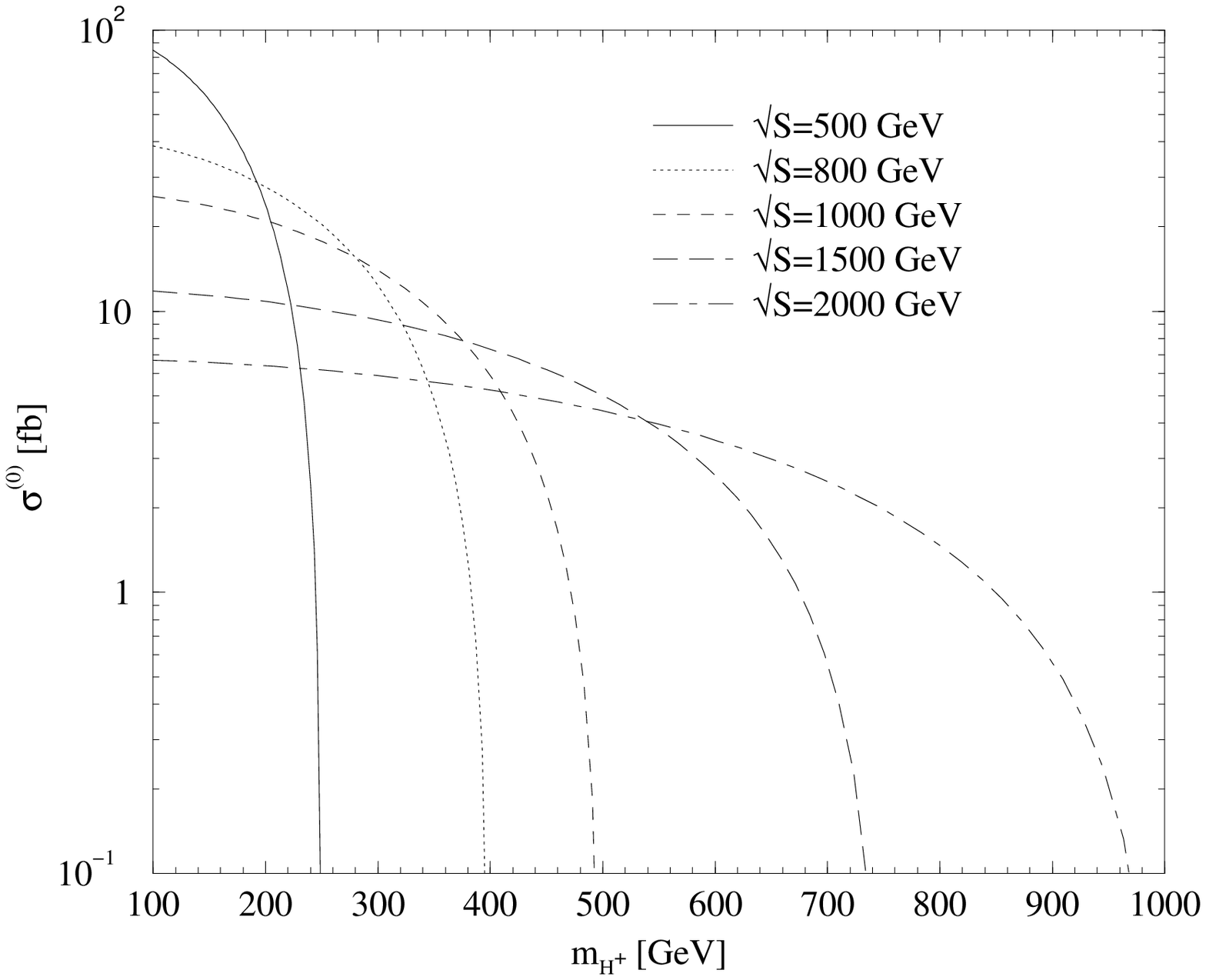}}
\end{center}
\caption{Lowest order cross-section for the process $e^+e^-\to H^+ H^-$ as a
  function of the charged Higgs boson mass, for different values of the center
  of mass energy $\sqrt{S}$.}\label{fig:tree}
\end{figure}
}

\newcommand{\figthdm}{
\begin{figure}[tbp]
\begin{center}
\resizebox{9.5cm}{!}{\includegraphics{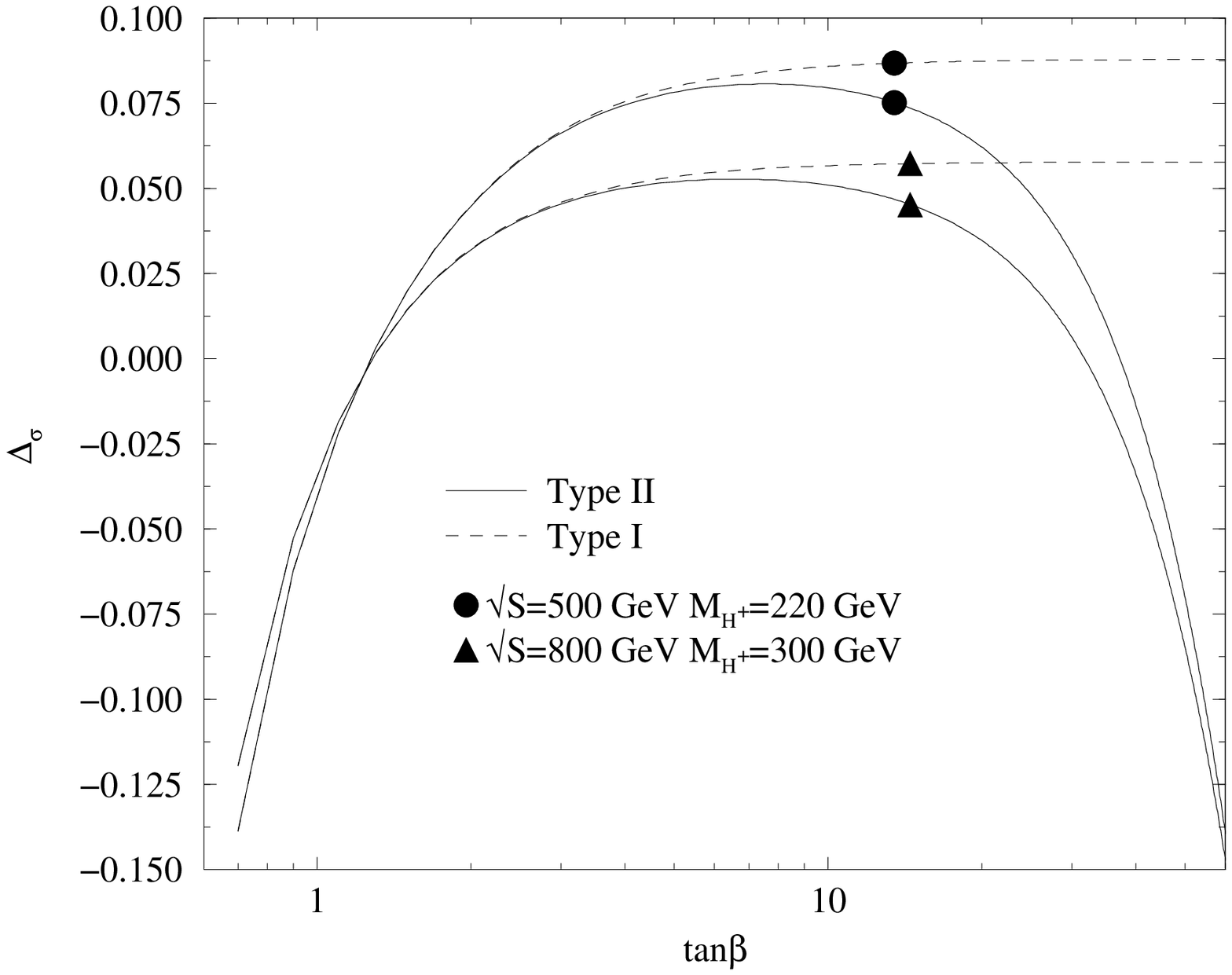}}
\end{center}
\caption{One-loop weak radiative corrections to $\sigma(e^+e^-\to H^+H^-)$ for the
  2HDM of type~I and~II as a function of \tb. The other parameters are:
  $\mh=150 \GeV$, $\mH=500\GeV$, $\mA=\mHp-20\GeV$, the scalar mixing angle
  $\alpha=\pi/2-\beta$, and the parameter $\lambda_5= 2 \pi \alpha_{em}
  \mH^2/(\ssw \mw^2)$.}
\label{fig:2hdm}
\end{figure}
}

\newcommand{\figmssmtb}{
\begin{figure}[tbp]
\begin{center}
\resizebox{10cm}{!}{\includegraphics{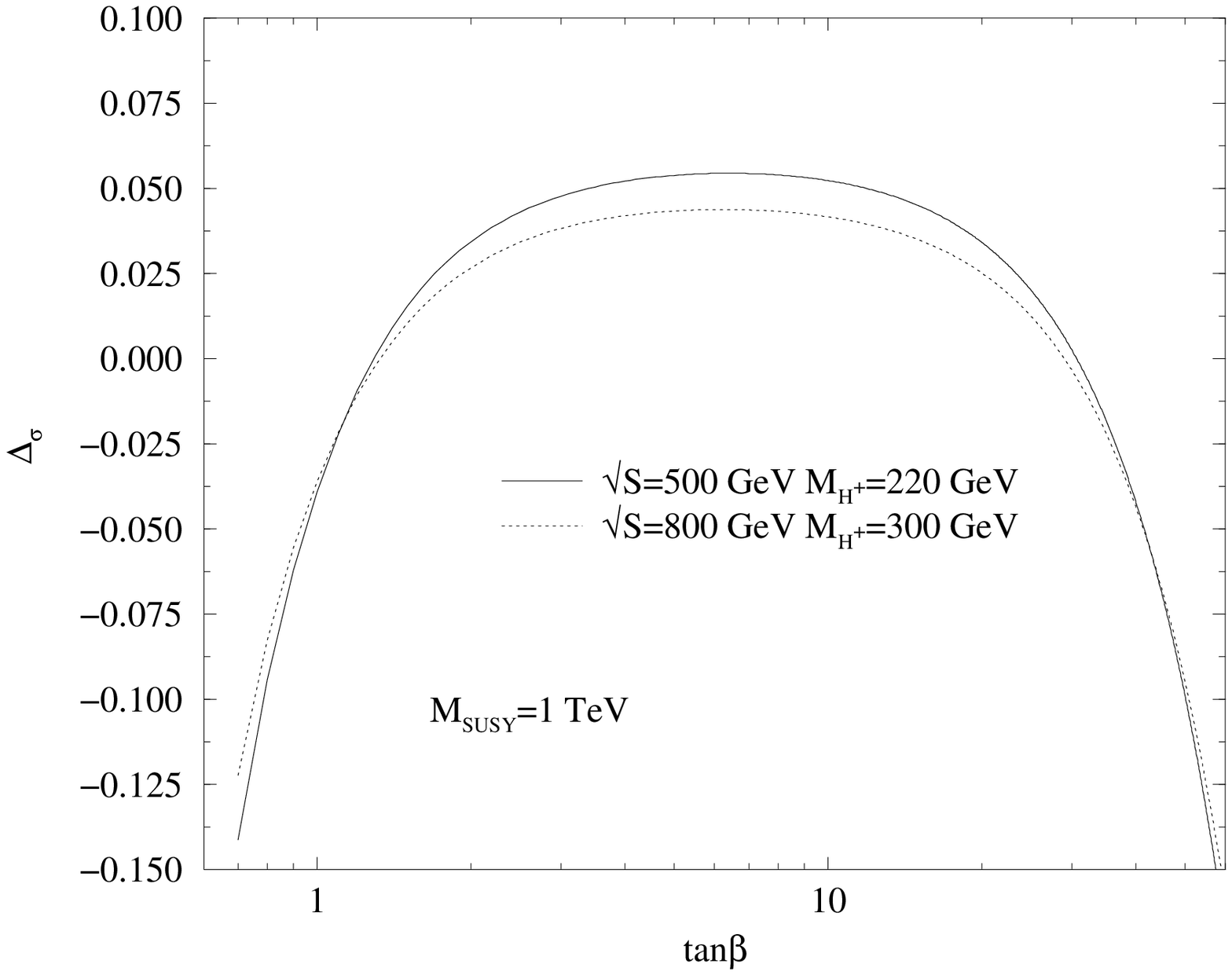}}
\end{center}
\caption{One-loop weak corrections to  $\sigma(e^+e^-\to H^+H^-)$ for the
  MSSM as a function of \tb. The
  SUSY parameters are taken to be $\mu=m_{\tilde q_L}=m_{\tilde q_R}=A_q=M_1=M_2=1 \TeV$.}
\label{fig:mssmtb}
\end{figure}
}

\newcommand{\figmssmmsb}{
\begin{figure}[tbp]
\begin{center}
\resizebox{10cm}{!}{\includegraphics{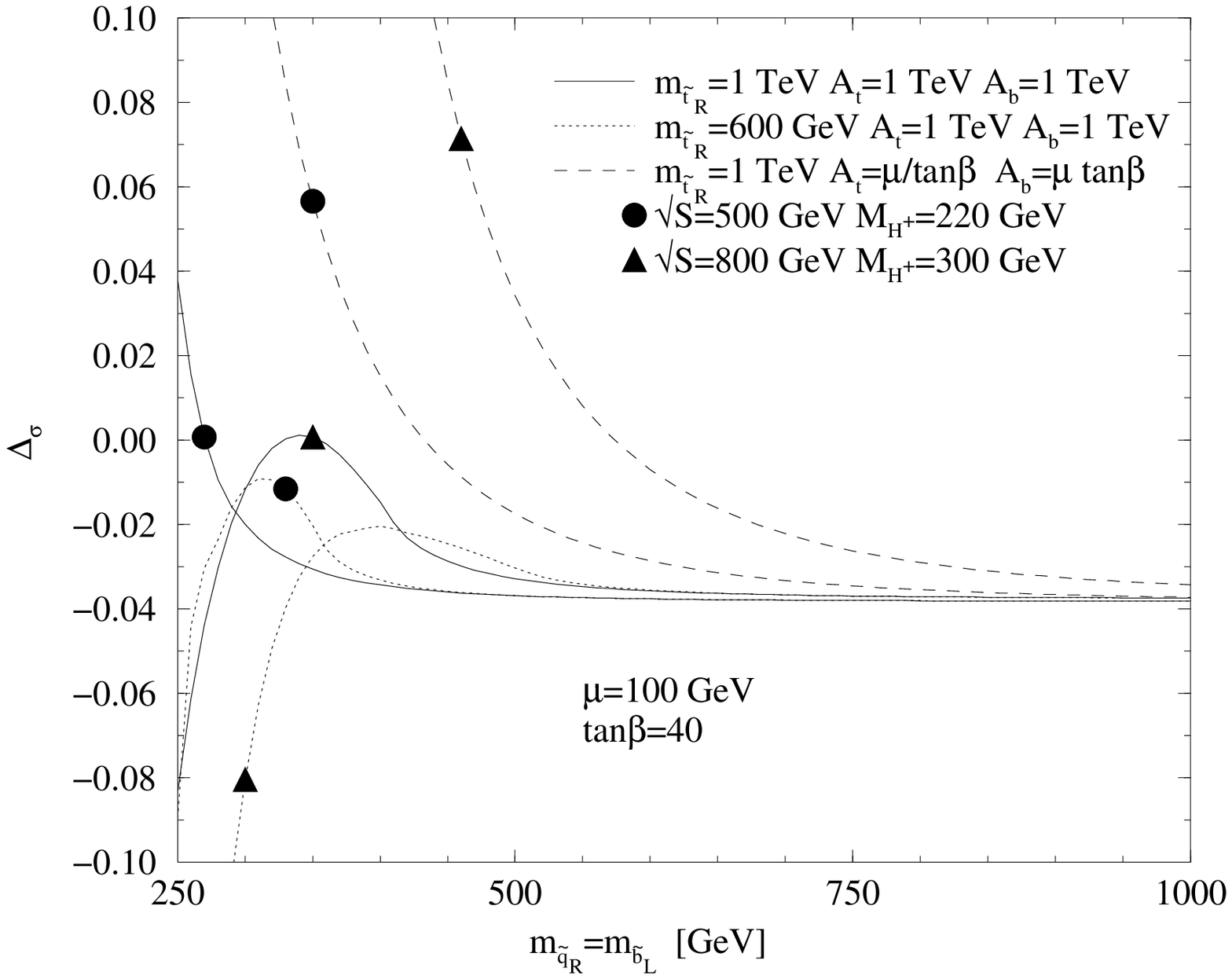}}
\end{center}
\caption{One-loop weak corrections to $\sigma(e^+e^-\to H^+H^-)$ in the
  MSSM as a function of the bottom squark mass parameter. The   other
  parameters are 
  given in the frame.}
\label{fig:mssmmsb}
\end{figure}
}

\newcommand{\figAfbthdm}{
\begin{figure}[tbp]
\begin{center}
\begin{tabular}{c}
\resizebox{10cm}{!}{\includegraphics{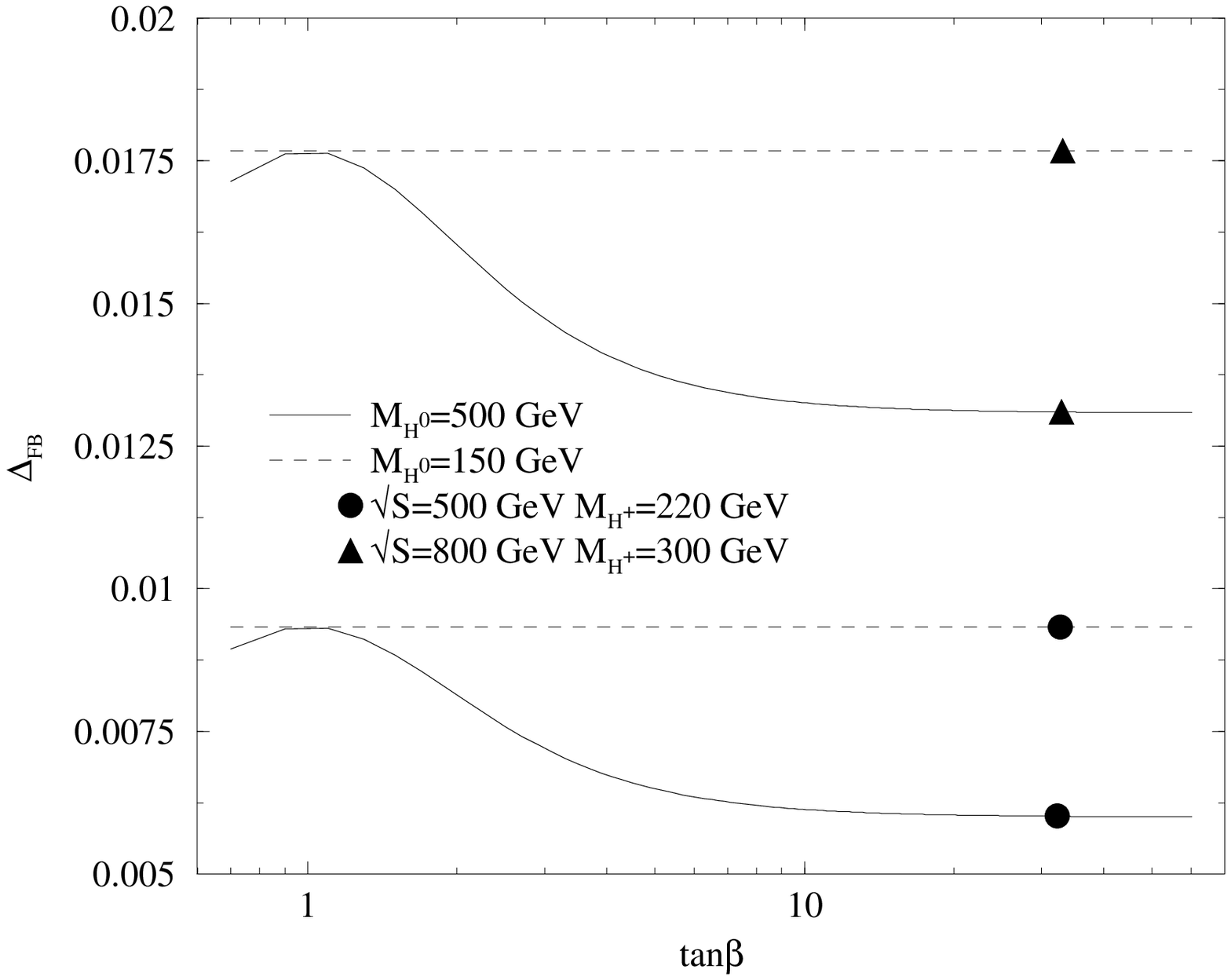}}\\
(a)\\
~\\
\resizebox{10cm}{!}{\includegraphics{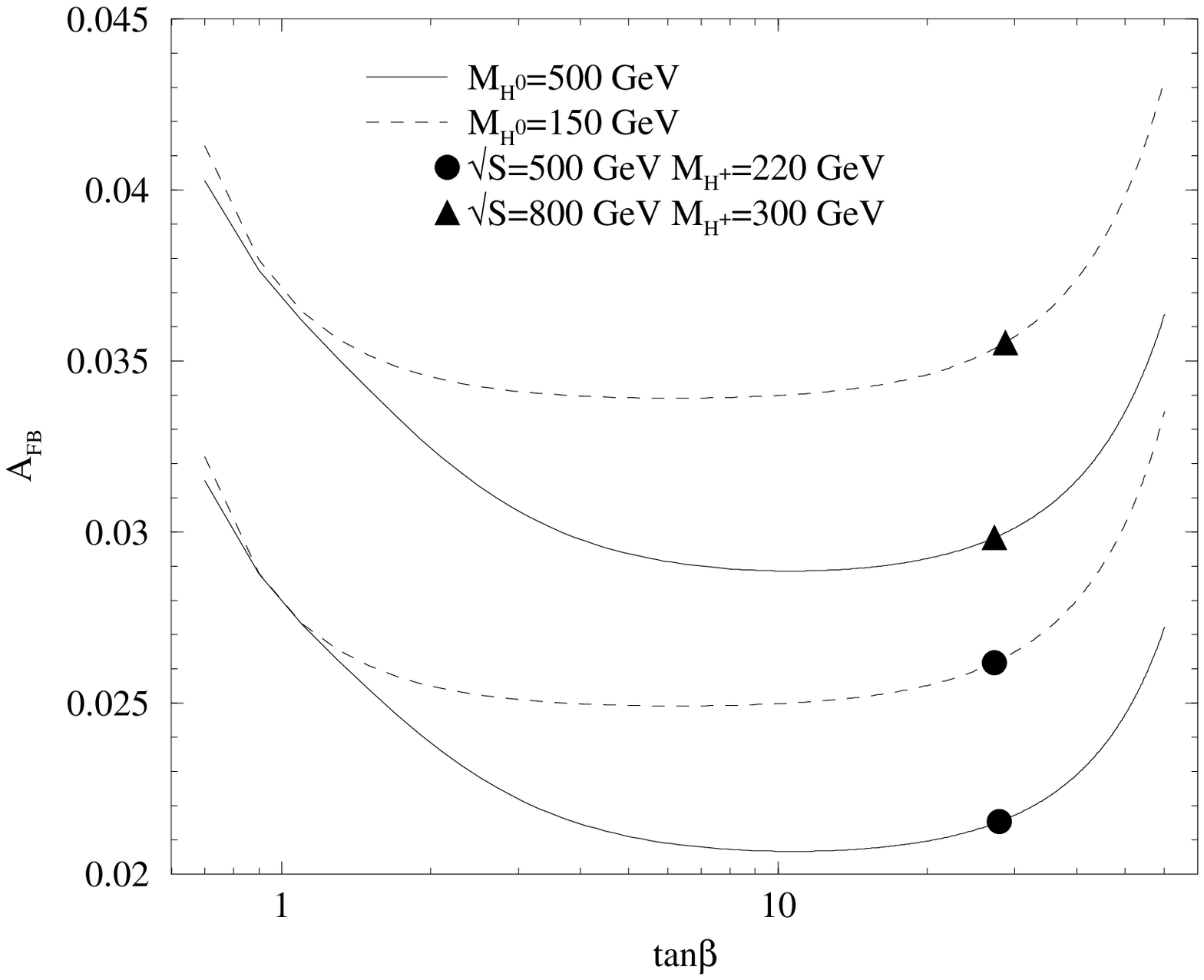}}\\
(b)
\end{tabular}
\end{center}
\caption{\textbf{(a)} Weak contributions to 
  $\Delta_{FB}$ eq.~(\ref{eq:Afb}), and \textbf{(b)}
  the total forward-backward asymmetry $\afb$ for the process $e^+e^-\to
  H^+H^-$ in the type II 2HDM as a function of \tb\  for different values of $\mH$
  as 
  given in the frame. The other parameters are  fixed as in Fig.~\ref{fig:2hdm}.}
\label{fig:afb2hdm}
\end{figure}
}

\newcommand{\figqed}{
\begin{figure}[tbp]
\begin{center}
  \resizebox{10cm}{!}{\includegraphics{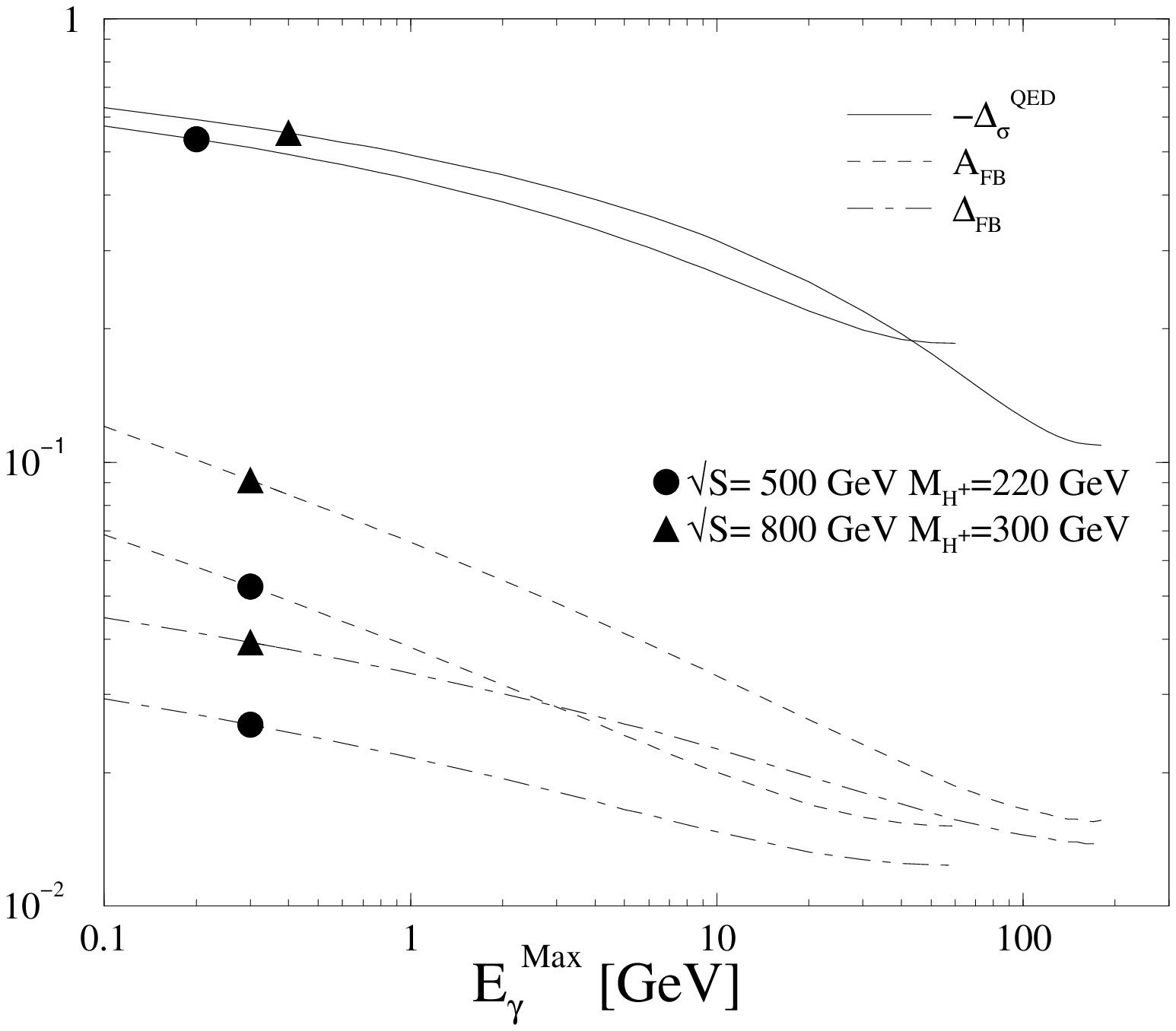}}
\end{center}
\caption{Photonic (QED) corrections to the total cross-section $\sigma(e^+e^-\to
  H^+H^-)$ (with reversed sign) and the forward-backward asymmetry  as a function of the
  photon energy cutoff.} 
\label{fig:qed}
\end{figure}}

\newcommand{\arra}[2]{$\begin{array}{c} #1 \\ #2 \end{array}$}

\newcommand{\tableAfb}{
\begin{table}
\begin{center}
\begin{tabular}{|c|c|c|c||c|c|c|}
\hline
\tb &  \arra{M_1}{{[}\GeV]}  & \arra{m_{\tilde{l}_L}}{{[}\GeV]} & \arra{m_{\tilde{e}_R}}{{[}\GeV]} & 
\arra{\Delta_\sigma}{{[}\%]} & \arra{\Delta_{FB}}{{[}\%]} & \arra{\afb}{{[}\%]}\\
\hline\hline
\multicolumn{7}{|l|}{set A: $\sqrt{S}=500\GeV$, $\mHp=220\GeV$}\\
\hline
40 &  500 & 1000 & 1000 & -4.160 &  0.874 & 2.728 \\
\hline
2 &  1000&  100 & 100  & 3.414 & 0.826 & 2.428\\
\hline\hline
\multicolumn{7}{|l|}{set B: $\sqrt{S}=800\GeV$, $\mHp=300\GeV$}\\
\hline
40 & 500 & 1000 & 1000  & -4.290 & 1.567& 3.491\\
\hline
2 &  1000&  100 & 100 & 2.752 & 1.530 & 3.182\\
\hline
\end{tabular}
\end{center}
\caption{The weak corrections to the integrated cross-section ($\Delta_\sigma$), the
  weak contributions to $\Delta_{FB}$ eq.~(\ref{eq:Afb}), and the total
  forward-backward asymmetry $\afb$ for the MSSM, for given parameters. Other
  parameters as in Fig.~\ref{fig:mssmtb}.} 
\label{tab:Afb}
\end{table}
}

\newcommand{\tablefinal}{
\begin{table}
\begin{center}
\begin{tabular}{|c|c|c|c|c|c|}
\hline
&$\Delta_\sigma$ & $\sigma\,[{\rm fb }]$ & $\afb$& $N$/year & $N^F-N^B$/year\\ \hline\hline
QED & $-18\%$ &$10$& $1.5\%$ &  5000 & 75 \\ \hline
2HDM & $-35\%$ to $-10\%$ & 8 to 11 & $2\%$ to $3.4\%$ & 4000 to 5500 & 110 to 135 \\ 
\hline
\end{tabular}
\end{center}
\caption{Total corrections ($\Delta_\sigma$), integrated cross-section ($\sigma$),
  forward-backward asymmetry ($\afb$), number of total expected events ($N$) and
  excess  of events in the forward direction ($N^F-N^B$), for the 2HDM of type II
  (set A: $\sqrt{S}=500\GeV$, $\mHp=220\GeV$), asuming an integrated luminosity
  of $500\,{\rm fb}^{-1}$/year.}\label{tab:final}
\end{table}
}

\begin{document}
\thispagestyle{empty}
\hfill \parbox{4cm}{KA-TP-19-1999\\
hep-ph/9911452\\
November 1999}

\vskip 2cm
\begin{center}
{\Large {\bf Radiative corrections to  pair production of charged Higgs
  bosons at
  TESLA\footnote{Talk presented by J. Guasch at the Vth workshop in the 2nd
  ECFA/DESY Study 
  on Physics and Detectors for a Linear Electron-Positron Collider,  Obernai
  (France) 16-19th October, 1999.} }}
\vskip 8mm

{\large Jaume Guasch, Wolfgang Hollik, Arnd Kraft}

\medskip
{\sl Institut f{\"u}r Theoretische Physik, Universit{\"a}t Karlsruhe,}

{\sl  D-76128 Karlsruhe, Germany}
\end{center}

\bigskip

\begin{center}
{\bf ABSTRACT}
\end{center}

\begin{quotation}
\noindent 
Charged Higgs particles are a common feature of many extensions of the Standard
Model. While its existence with masses up to the TeV scale can be probed at
the Tevatron Run II and LHC hadron colliders, a precise determination of its
properties would have to wait for a high energy $e^+e^-$ linear collider. We have
computed the complete one-loop electroweak corrections to the cross-section for
charged Higgs bosons pair production in $e^+e^-$ collisions, in the framework of
general Two Higgs Doublet Models, as well as the Minimal Supersymmetric Standard
Model. A study is presented for typical values of the model parameters, showing
the general behaviour of the corrections. 
\end{quotation}

\newpage
\setcounter{page}{1}
\section{Introduction}
During the next decades the Tevatron and LHC hadron colliders will explore the 
TeV energy region, performing tests of the Standard Model (SM) at energies
higher than ever before and searching for new physics phenomena. 
While these machines are well suited to discover possible new
particles with large masses 
(e.g. up to $2.5\TeV$ for the LHC~\cite{GianOber}), the next generation
of high energy and high luminosity $e^+e^-$ linear colliders, such as
TESLA, are the best environment to make a precise
determination of their properties. 
One of the major goals of TESLA will be to perform
precision measurements of the properties of heavy particles, 
like the top quark, reaching a level 
comparable to the LEP I determination of the $Z$ boson observables. 
In view of these prospects it is
necessary to account for radiative corrections in order to
have theoretical predictions with an uncertainty smaller than the 
experimental one.
In the case of processes involving non-standard particles, precision studies
are a crucial requirement for their accurate identification within
the various conceivable extensions of the SM.  

A common feature of many SM extensions is the 
augmentation of the Higgs sector, predicting
the presence of additional neutral Higgs bosons
and of charged Higgs particles, which are in the focus of this study. 
The simplest SM extension,
the so-called Two Higgs Doublet Model (2HDM)~\cite{HHG},   
includes an additional Higgs doublet in the scalar sector. 
From a theoretical point of view, the  supersymmetric
(SUSY) extensions of the SM are more appealing; 
the simplest version in this class of models is the Minimal
Supersymmetric Standard Model (MSSM)~\cite{MSSM}.

In the MSSM, the Higgs sector is determined at the tree-level by just two
parameters, which are conventionally chosen to be 
the ratio of the two vacuum expectation values,
$\tb=v_2/v_1$, and a mass which we take as the charged Higgs mass
$\mHp$. The relations between the 
various masses, however,
 get significant corrections at higher orders~\cite{Dabels}.

In the general 2HDM, all the 
masses and the  mixing
angle  $\alpha$ for the CP-even neutral Higgs bosons are independent 
parameters~\cite{HHG}; in addition we have an independent parameter which enters
the Higgs bosons self-couplings,
which we take to be $\lambda_5$ in the convention of~\cite{HHG}.
Indirect constraints on charged Higgs bosons are obtained from
present low energy data on the branching ratio $BR(b\to s\gamma)$
which does 
not favour low values of the charged Higgs mass in the 2HDM of 
type~II~\cite{Cleo};
the latest analysis indicates a value for $\mHp$ higher
than $165\GeV$~\cite{Borzubsg}.

If the charged Higgs particle is light enough, it
will be pair-produced at TESLA at appreciable rates. 
The detailed discussion of the production cross-section on the
basis of a complete one-loop calculation 
in the general 2HDM and the MSSM, 
incorporating also soft and hard photon bremsstrahlung,
is the content of this article.

\figtree

The first question to be addressed is the relevance 
of studying charged Higgs pair production.  
In Fig.~\ref{fig:tree} we have displayed the tree-level
cross-section 
$\sigma^{(0)}(e^+e^-\to H^+H^-)$
as a function of the charged Higgs mass for
different values of the center of mass energy $\sqrt{S}$.  
The typical value of this cross-section lies in the ballpark 
of several ${\rm fb}$, unless
the center of mass energy is very close to the production threshold.

For definiteness, 
we will concentrate our discussion in this article on two scenarios: 
the first one with an energy of
$\sqrt{S}=500 \GeV$ and a charged Higgs of $\mHp=220\GeV$ (set A), 
the second one with a
higher energy of $\sqrt{S}=800 \GeV$ 
and a charged Higgs mass of $\mHp=300\GeV$ (set B). 
A high luminosity of
$500\, {\rm fb}^{-1}/{\rm year}$ for the TESLA collider is assumed.
Under these conditions 
$\sigma^{(0)}(H^+H^-)= 11.8\,{\rm fb}$ [$12.3\,{\rm fb}$] for set A [B]
is obtained, 
which means the production of ca.\
5900 [6150] pairs of charged Higgs particles per year.
For set B a recent study~\cite{Kiiskinen} 
shows that the background to this process can be
reduced very efficiently, 
maintaining most of the signal events and allowing
for a precise determination of the mass of the charged Higgs particle.

There are also other processes where a charged Higgs boson could be produced
at TESLA. The associated production 
$e^+e^-\to H^{\pm} W^{\mp}$~\cite{zhu} is one-loop mediated
in lowest order, and thus its cross-section is smaller 
than that of $H^+H^-$ pair production, provided the latter channel is open. 
There exists also the associated
production with quarks of the third generation $H^+ \bar{t} b$, which suffers
of being a three-body production; but it could be important in certain regions
of the parameter space. Finally, if the charged Higgs is light enough, 
it could arise from primary top quarks decaying according to 
$t\to H^+ b$.
The non-SM radiative corrections to this process have been studied, 
and are
found to be very large, with important consequences for the charged Higgs
search in the region $\mHp<\mt$~\cite{tbh}.

The second question to be addressed is the size of the radiative corrections 
to $e^+e^-\to H^+H^-$. They can be expected to
be important because of the
potentially large couplings associated with the Higgs sector, to wit:
the Yukawa coupling of
the top quark, and 
of the bottom quark at large \tb\ for the type~II 2HDM; the
couplings involving squarks of the third generation in the MSSM; 
and the self-couplings
of the Higgs particles in the general 2HDM. 

The third  question 
to be addressed is the possibility to
discriminate between the different models by means of precise
measurements of the cross-section.
This point is closely related to
the previous one, since the 
detailed structure of the model enters only at the quantum level,
through the radiative corrections.

In this note we will restrict 
ourselves to the presentation of the main
results of the one-loop calculation; the complete expressions and a
comprehensive 
analysis will be presented elsewhere~\cite{Ourpaper}. 
The computation has been thoroughly checked,
invoking also 
a numerical check versus a  code
generated by the computer-algebra system \textit{FeynArts} and
\textit{FormCalc}~\cite{FAFC}.
The one-loop corrections
for this process were also discussed
in Ref.~\cite{frenchsb} for the fermion and
sfermion contributions, and in 
Refs.~\cite{frenchhiggs,frenchnew} for the Higgs, gauge and fermion
sector in the case of the 2HDM of type II and also of the MSSM Higgs sector. 
Our results are in good agreement with those of
Ref.~\cite{frenchnew}. 
In the present work we have also computed the
corrections for the case of the type I 2HDM and the
contributions from the chargino-neutralino sector in the MSSM.
Moreover, we present an analysis of the forward-backward asymmetry,
including also the complete $O(\alpha)$ QED corrections with
hard photons.

\section{One-loop corrections to $\sigma(e^+e^-\to H^+H^-)$}

The lowest-order differential cross-section for $e^+e^-\to H^+H^-$
is determined exclusively by the gauge couplings and by the
mass $\mHp$, which enters only  the phase space.
Moreover, the angular distribution is symmetric in the production
angle $\theta$. Through the loop contributions, the cross-section
becomes dependent on the details of the Higgs sector and, in addition,
of the SUSY particles in the case of a supersymmetric model.
Specifically, also an angular asymmetry is induced, giving rise to
a forward-backward or charge asymmetry.  

\subsection{QED corrections}
As often  done in the discussion of one-loop corrections, we first
separate the 
subclass of the  QED (photonic) corrections. The contributions of this
class are universal
in the sense that they only depend on the charge of the particles but
not on the details of the underlying model. 
They are in general numerically important, and it is necessary
to have them under control if one wants to observe 
the small effects resulting from the residual non-QED one-loop 
contributions with their model-specific informations. 

We have computed the complete
${\cal O}(\alpha)$ QED corrections, which arise from the exchange of
virtual photons and from real photon bremsstrahlung. 
Both the soft and the hard photon emission are included.
The bulk of the QED corrections is 
given by initial state radiation (ISR), which 
lowers the effective center-of-mass energy available
for the annihilation process. 
These corrections are typically quite large; one therefore has to 
take into account also higher orders with a resummation of leading
terms.
For our calculation we have used 
the resummed leading 
ISR corrections as given by the structure function
method~(see e.g.~\cite{LEPIIyellow}).

The interference between initial and final state radiation 
is of special interest since it is responsible, together with 
the box diagrams containing at least one virtual photon,    
for a QED-induced charge asymmetry, which has to be well separated
from the charge asymmetry caused by the model-specific non-QED
box diagrams.

\figqed

In Fig.~\ref{fig:qed} we display the QED corrections to the 
integrated cross-section (with reversed sign)
as well as the QED contributions to the forward-backward
asymmetry $\afb$ and 
to the quantity $\Delta_{FB}$ in eq.~(\ref{eq:Afb}),  as a
function of the cut-off to the photon energy
assuming that photons with energy
$E\leq E_{\gamma}^{Max}$ are not resolved. 
The relative correction $\Delta_\sigma$ 
is defined in the following way 
\begin{equation}
\sigma=\sigma^{(0)}+\delta\sigma=\sigma^{(0)}(1+\Delta_\sigma)\,\,,
\label{eq:dsigdef}
\end{equation}
where $\sigma$ is the integrated cross-section 
including radiative corrections and
$\sigma^{(0)}$ is the tree-level prediction. 
We see that for a more inclusive cross-section, 
allowing for photons with higher energies, 
the QED corrections decrease.
For the choice of parameters as in Fig.~\ref{fig:qed} 
the QED corrections
to the total cross-section are negative, ranging from $-60\%$ to $-20\%$
[$-10\%$] for set A [B]. 
The asymmetry is discussed in section 3.

\subsection{Weak Corrections}

The weak corrections form the complementary subclass of the non-QED
one-loop contributions, with model-specific information beyond that in the
tree-level amplitude.
In Figs.~\ref{fig:2hdm}-\ref{fig:mssmmsb} 
we present a summary of the numerical
results found for the pure weak corrections, 
pointing out their typical behaviour. For 
the interpretation of the analysis it is useful to separate the different
contributions arising from the various particles in the loops,
as well as to distinguish 
between initial state (IS) corrections (the $e^+e^-$ vertex corrections), 
final state (FS)
corrections (the $H^+H^-$ vertex corrections), 
self-energies of the intermediate photon and $Z$
boson, and box diagrams where two internal lines connect 
the external $e^+e^-$ legs to the external
$H^+H^-$ legs. 
One should keep in mind, however, that for loops with gauge bosons the
separation between the different topologies is in general 
gauge dependent.

We  classify the weak corrections as follows:
\begin{itemize}
\item The first set of diagrams is formed by loops which contain gauge bosons,
  together with electrons, neutrinos, and Higgs particles. These particles
  contribute to the IS, FS, self-energies, and box corrections.
  Their contributions are sizeable (see below), and they are
  specially interesting in the parameter range of $\tb=2-20$. 
\item Vertex one-loop diagrams with Higgs boson exchange only contribute to the
  FS corrections; in the IS they are negligible. 
  In these kind of diagrams  the potentially large Higgs
  self-couplings occur in the  general 2HDM, whereas in the MSSM
  the Higgs self-couplings are always small.
\item The loops with top and bottom quarks 
   contribute to the FS corrections through the
  Yukawa couplings, which are given by~\cite{HHG}
  \begin{equation}
  \lambda_{t}\equiv{\frac{m_t}{v_2}}
    \sim \frac{m_t}{\sin\beta}
   \;\;\;\;\;,\;\;\;\;\;
   \lambda_{b}^{\{I,\,II\}}\equiv{\frac{m_b}{\{v_2,v_1\}}}
   \sim \frac{m_{b}}{\{\sin{\beta},\,\cos\beta\}}\,\,, \label{eq:Yukawas}
  \end{equation}
  where the expression for $\lambda_b$ depends on the type of the 2HDM. We see
  that these couplings  
  can be important either at low and (for the type~II 2HDM) 
  large values of \tb. 
  In the MSSM, only type II is realized.
\item In the MSSM we have in addition the corrections due to squarks, 
  with the most
  important ones from   
  the third generation squarks ($\stopp,\sbottom$). In
  this case not only the Yukawa-type couplings~(\ref{eq:Yukawas}) are large, 
  but also the trilinear 
  squark-squark-Higgs boson couplings can be quite huge, 
  leading to large corrections.
\item Finally we have the diagrams with loops containing  
  all the plethora of chargino, neutralino,
  and selectron-sneutrino. These particles contribute to 
  all kind of topologies. 
  As we have found, their contribution to the total
  cross-section is generally rather small, at the level of a 
  few per mille; however, they do contribute to
  the forward-backward asymmetry.
\end{itemize}

\figthdm

We start our numerical analysis with the case of the general 2HDM. In
Fig.~\ref{fig:2hdm} we present the evolution of the corrections with \tb\  for
fixed typical values of the other parameters. It should be stated that in all
our numerical analysis we have respected present constrains on the parameter
space, taking care that the additional contributions to the $\rho$ parameter do
not surpass present experimental errors. 

The free parameter $\lambda_5$ in the Higgs bosons self-couplings is a general
feature of the unconstrained 2HDM. For $0.7<\tb<2$ it modifies the cross-section
not more than $\pm10\%$, even for rather large values $\lambda_5<20$. For large
\tb, however, $\lambda_5$ enters the loop diagrams in the combination
$\sim\lambda_5\tb$. This gives rise to large positive corrections, which would
be in conflict with perturbative consistency if we let $\lambda_5$ vary
freely. In the following we restrict our discussion to small values of $\lambda_5$. 

From Fig.~\ref{fig:2hdm} we can see 
that for low \tb\ values both of the 2HDMs  give the same
predictions. In this region the radiative corrections are dominated by the top
quark Yukawa coupling~(\ref{eq:Yukawas}) 
and can amount to $-15\%$. The
size of the corrections decreases 
with the top quark Yukawa coupling, and a flat
region exists for the intermediate
range of $\tb=2-10$. In this region the radiative
corrections are driven by the gauge- and Higgs self-couplings. 
For our chosen set of
parameters they amount to a $\sim7\%$ positive
correction. For larger energies or lower charged Higgs boson masses the
corrections can be slightly negative~\cite{Ourpaper}. For the type~II 2HDM at large
\tb,  the bottom quark 
Yukawa coupling~(\ref{eq:Yukawas}) enters the game, 
driving again towards negative 
corrections for large \tb. For the type~I 2HDM, on the other hand, the
corrections reach an asymptotic value.

\figmssmtb
\figmssmmsb

In Fig.~\ref{fig:mssmtb} we have plotted the weak corrections for the case of
the MSSM, 
assuming a SUSY mass spectrum in the ballpark of $1\TeV$. The
corrections show a behaviour similar to that of a type~II 2HDM. If the
SUSY spectrum is lighter, however, the situation changes. 
Fig.~\ref{fig:mssmmsb} shows
the corrections as a function of the sbottom mass parameter, for different choices of
the mixing parameters. Indeed, when the squark masses get low
enough ($m_{\tilde q}\lsim 500\GeV$) 
they give positive contributions to the
corrections, 
which can be sizeable even when they are too heavy to be directly produced at
TESLA. 
Lighter squark masses can make the corrections negative, 
thereby yielding  much 
larger values than the 2HDM case for the same value of \tb. This is the 
only potentially large loop effect from SUSY particles,
since the contributions from the
chargino-neutralino sector are 
comparatively tiny (below the $1\%$ level).

\section{The forward-backward asymmetry}

In principle the forward-backward asymmetry seems to be an appropriate 
observable for testing the type of electroweak model for the charged
Higgs bosons.
In the case of the MSSM there are much more
box-type diagrams (with virtual charginos and neutralinos) 
that contribute to the asymmetry, 
and one might expect that, 
although their effect on the total cross-section is negligible, 
they could amount to an observable contribution to the asymmetry. 

The forward-backward asymmetry is defined as
\begin{equation}
\afb=\frac{\sigma^F-\sigma^B}{\sigma}=\Delta_{FB}
\frac{1}{1+\Delta_\sigma}\,\,.
\label{eq:Afb}
\end{equation}
$\Delta_{FB}$ denotes the antisymmetric part of the cross-section
normalized to the tree-level cross-section,
\begin{equation}
\Delta_{FB} = \frac{\sigma^F - \sigma^B}{\sigma^{(0)}} ,
\end{equation} 
and $\Delta_{\sigma}$ is the relative correction for the integrated 
cross-section, see eq.~(\ref{eq:dsigdef}).
Formally
the difference between $\afb$ and $\Delta_{FB}$ is of higher order;
however, owing to the presence of 
large corrections to the total cross-section, it is numerically significant.
Note that $\Delta_\sigma$ in eq.~(\ref{eq:Afb}) should contain all the
QED and weak one-loop contributions. 
Although the physical quantity to study is
$\afb$, the quantity $\Delta_{FB}$ can be written as the sum of the various
contributions, and it is thus useful to make comparisons between the
different sources of the asymmetry. 

\figAfbthdm

In Fig.~\ref{fig:qed} 
we have already presented the part of $\Delta_{FB}$  resulting
from the pure photonic contributions 
and the asymmetry $\afb$ including only QED effects, as a function
of the photon energy cut-off.
The QED induced $\afb$ is rather large, reaching from
$\afb=6\%$ [$10\%$] to $\sim1.5\%$ for set A [B]. 
A large part, however,  is due to the large
negative corrections to the integrated cross-section. 
The quantity $\Delta_{FB}$ is independent
of this normalization effect. 
As can be seen in Fig.~\ref{fig:qed}, the QED contributions 
to $\Delta_{FB}$ are between $3\%$  [$4\%$] for a very restrictive
cut to the photon energy, and decrease 
to $\sim 1\%$ for a totally inclusive treatment of the phase space. 
Since for a totally inclusive
measurement the QED effects are largely reduced (both, in the asymmetry and 
the total cross-section)
the purely weak effects could be accessible more easily.
For definiteness we shall assume in the following discussion always 
the inclusive case, which corresponds to the fully integrated 
phase space of the radiated photon.

We now turn to the weak contributions. In the 2HDM, the
quantity $\Delta_{FB}$ depends only mildly on 
the parameters and on the type of model. 
The dependence on the model parameters essentially enters through
the normalization factor $\Delta_\sigma$. Fig.~\ref{fig:afb2hdm} 
shows the part of $\Delta_{FB}$ resulting from the weak-loop diagrams
(to be added to
the QED one of Fig.~\ref{fig:qed}) 
and the total forward-backward asymmetry
$\afb$, including the fully inclusive QED effects,
for the general 2HDM. 
In Fig.~\ref{fig:afb2hdm}a we see that the weak part of $\Delta_{FB}$ 
amounts to about $1\%$ [$1.75\%$] for set A [B], 
which is of the same order as the
QED-induced part in each case (Fig~\ref{fig:qed}). This means that the
total forward-backward asymmetry would be largely enhanced 
with respect to the QED expectations. As can be seen  
in Fig.~\ref{fig:afb2hdm}b,  $\afb$ can
be  twice the totally inclusive asymmetry of Fig.~\ref{fig:qed};
even at its minimum it is significatively larger.

\tableAfb

The additional MSSM contributions 
owing to charginos and neutralinos are smaller;
their effect is typically  a $10\%$ reduction of 
$\Delta_{FB}$ from virtual gauge and Higgs bosons.
To make this more illustrative, we present in
table~\ref{tab:Afb} the weak-loop part of
$\Delta_{FB}$ in the MSSM, and the total asymmetry $\afb$ for various sets
of model parameters. 
The parameters used in table~\ref{tab:Afb} are representative
for the extreme values  $\afb$ can take in the MSSM 
(for the given values of
$\sqrt{S}$ and $\mHp$). We can see that the variation of $\Delta_{FB}$ is
rather small, and the main variation of $\afb$ is again 
mainly due to the corrections
to the normalization factor.

In general, $\Delta_{FB}$ (and so $\afb$) as a function of $S$, increases with the velocity of
the produced 
charged Higgs boson $\beta=\sqrt{1-4 \mHp^2/S}$.

\section{Conclusions}
We have presented a full ${\cal O}(\alpha)$ computation of the radiative
corrections to the production cross-section for
$\sigma(e^+e^-\to H^+H^-)$ together with a discussion for values of
the charged Higgs mass relevant to the TESLA $e^+e^-$ linear collider. 
The computation was done in the general framework 
of the type~I and type~II 2HDM, as well as in the MSSM.

The one-loop corrections owing to virtual 
gauge and Higgs particles are typically of the size  $\sim 5\%$, but they can be
much larger (positive) for large values of $\lambda_5$ in the general 2HDM. 
The ones from top and bottom quarks are negative; they can be as
large as $\sim-20\%$  in the low \tb\ region,
and for the type~II 2HDM and the MSSM also in the region of large \tb.  

The main effect from SUSY particles to the total cross-sections 
originates from the squarks of the third generation, yielding
large  corrections provided $m_{\tilde q}\lsim 500\GeV$. The
chargino-neutralino contributions are of the order of a few per mille.

The forward-backward asymmetry
produced by weak loop effects ($\Delta_{FB}$) is found
to be around  the 
$1\%$ [$2\%$] level in the general 2HDM for set A [B]. The additional SUSY
contributions to $\Delta_{FB}$, mediated by 
charginos and neutralinos, are smaller, producing a decrease of $10\%$ 
with respect to the 2HDM expectations.

The QED corrections from real and virtual photons
are rather large: $-60\%$ to $-20\%$ [$-10\%$] for the integrated
cross-sections,
and $6\%$ [$10\%$] to $1.5\%$ for the asymmetries. 

\tablefinal

With the QED corrections under control 
it should be feasible to
detect the genuine weak quantum contributions to the total cross-section. 
With a total
inclusive measurement with respect to the photons, 
the QED effects can be largely diminished; 
in this case the QED contribution to $\Delta_{FB}$ (eq.~(\ref{eq:Afb}))
is of the same order as the weak loop contribution.
Hence, the forward-backward asymmetry is
significantly increased (even doubled) compared to the QED expectations.  
In table~\ref{tab:final}
we present the expected number of events 
taking into account only the QED corrections (first line) and the complete
one-loop corrections for the 2HDM case (second line), 
for an integrated luminosity of
$500\,{\rm fb}^{-1}$ and for the parameter set A.
The numbers show that the effects in the asymmetry should be detectable. 
On the other hand, to disentangle the effects resulting from
charginos (see table~\ref{tab:Afb}) from 
the Higgs and gauge boson contributions, 
will require precision measurements below the 1\% level.

The size of the radiative corrections depends slightly on the mass of 
$\hplus$ and the energy  $\sqrt{S}$, with similar 
conclusions as given above.
Two examples are contained  in the numerical analysis of this
article; a comprehensive description will be given in~\cite{Ourpaper}.

\section*{Acknowledgments}
We would like to thank Thomas Hahn and Christian Schappacher for the 
computer-algebra-generated check. 
We are also thankful to Abdesslam Arhrib for numerical
comparisons. 
This work has been partially supported by the Deutsche Forschungsgemeinschaft.


\begin{thebibliography}{99}
\bibitem{GianOber} F.~Gianotti, talk  at the Vth workshop in the 2nd
  ECFA/DESY Study 
  on Physics and Detectors for a Linear Electron-Positron Collider,  Obernai
  (France) 16-19th October, 1999.

\bibitem{HHG} J.~Gunion, H.E.~Haber, G.L.~Kane, S.~Dawson, {\it The
Higgs Hunter's Guide} (Addison-Wesley, Menlo-Park, 1990); \textit{erratum} \texttt{hep-ph/9302272}.

\bibitem{MSSM} H.~Nilles,\, {\it Phys. Rep.} {\bf 110} (1984) 1;\\
 H.~Haber, G.~Kane, {\it Phys. Rep.} {\bf 117} (1985) 75;\\
 A.~Lahanas, D.~Nanopoulos, {\it Phys. Rep.} {\bf 145} (1987) 1;\\
 See also the exhaustive reprint collection {\bf Supersymmetry}
(2 vols.), ed. S.~Ferrara (North Holland/World Scientific, Singapore, 1987).




\bibitem{Dabels}
H.E.~Haber, R.~Hempfling,
\textit{Phys.\ Rev.\ Lett.\ }{\bf 66} (1991) 1815;\\
Y.~Okada, M.~Yamaguchi,  T.~Yanagida,
\textit{Prog.\ Theor.\ Phys.\ }{\bf 85} (1991) 1;\\
J.~Ellis, G.~Ridolfi, F.~Zwirner,
\textit{Phys.\ Lett.\ }{\bf B257} (1991) 83;
\textit{ibid.} {\bf B262} (1991) 477;\\
R.~Barbieri, M.~Frigeni,
\textit{Phys.\ Lett.\ }{\bf B258} (1991) 395;\\
  P.~Chankowski, S.~Pokorski, J.~Rosiek,
  \textit{Nucl. Phys.} \textbf{B 423} (1994) 437; \\
  A.~Dabelstein,  \textit{Z. Phys.} \textbf{C 67 }(1995) 495;
  \textit{Nucl. Phys.} \textbf{B 456} (1995) 25; \\
J.A.~Bagger, K.~Matchev, D.M.~Pierce, R.~Zhang,
\textit{Nucl.\ Phys.\ }{\bf B491} (1997) 3;\\
M.~Carena, M.~Quir{\'o}s, C.~Wagner, \textit{Nucl. Phys.} \textbf{B 461} (1996) 407; \\
  H.~Haber,  R.~Hempfling, A.~Hoang, \textit{Z. Phys}. \textbf{C 75} (1997) 539;  \\
  S.~Heinemeyer, W.~Hollik, G.~Weiglein, \textit{Phys. Lett.} \textbf{B 440}
  (1998) 296; \textit{Phys. Rev. }{\bf D58} (1998) 091701; \textit{Eur.\ Phys.\ J.\ }{\bf C9} (1999) 343.

\bibitem{Cleo} M.~S. Alam \textit{et~al.} (CLEO Collaboration), 
\textit{Phys. Rev. Lett. }\textbf{ 74} (1995) 2885;\\
J.~Alexander, 
in \textit{Proceedings of the 29th International Conference on High-Energy
  Physics}, Vancouver 1998, eds. A.~Astbury, D.~Axen, J.~Robinson,
(World Scientific, Singapore, 1999), p.~129;\\
S.~Ahmed {\it et al.}
(CLEO Collaboration),
hep-ex/9908022.


\bibitem{Borzubsg}  F.~Borzumati, C.~Greub, \textit{Phys. Rev.} \textbf{D58}
  (1998) 074004; {\it ibid.} \textbf{D59} (1999) 057501.

\bibitem{Kiiskinen} A.~Kiiskinen, talk at the Vth workshop in the 2nd ECFA/DESY Study
  on Physics and Detectors for a Linear Electron-Positron Collider,  Obernai
  (France) 16-19th October, 1999.


\bibitem{zhu} S.H.~Zhu, \texttt{hep-ph/9901221};\\
A.~Arhrib, C.~Capdequi Peyran{\`e}re, W.~Hollik, G.~Moultaka, in preparation.

\bibitem{tbh} J.~Guasch, talk  at the  IVth workshop in the 2nd
  ECFA/DESY Study on Physics and Detectors for a Linear Electron-Positron
  Collider, Oxford  20-23 March 1999; \\
C.~Li, B.~Hu, J.~Yang,
\textit{Phys.\ Rev.\ }{\bf D47} (1993) 2865, \textit{erratum ibid.} \textbf{D48} (1993) 3410;\\
J.~Guasch, R.A.~Jim{\'e}nez, J.~Sol{\`a},
  \textit{Phys. Lett. }\textbf{ B360} (1995) 47;\\
J.A.~Coarasa,\,D.~Garcia,\,J.~Guasch,\,R.A.~Jim{\'e}nez,\,J.~Sol{\`a},\,\textit{Eur.\,Phys.\,J.}\,\textbf{C2} (1998) 373;\\
J.~Guasch, J.~Sol{\`a},  \textit{Phys. Lett. }\textbf{B416} (1998) 353;\\
J.~A. Coarasa, J.~Guasch, W.~Hollik, J.~Sol{\`a}, \textit{Phys. Lett. }\textbf{B442} (1998) 326. 

\bibitem{Ourpaper} J.~Guasch, W.~Hollik, A.~Kraft, preprint KA-TP in preparation.

\bibitem{FAFC}
J.~K{\"u}blbeck, M.~B{\"o}hm, A.~Denner,
\textit{Comput.\ Phys.\ Commun.\ }{\bf 60} (1990) 165;\\
T.~Hahn, \textit{FeynArts 2.2 user's guide}, \texttt{http://www.feynarts.de}; \\
T.~Hahn, \textit{FormCalc and LoopTools user's guide}, \\ \texttt{http://www-itp.physik.uni-karlsruhe.de/formcalc}.



\bibitem{frenchsb}
A.~Arhrib, M.~Capdequi Peyran{\`e}re, G.~Moultaka,
\textit{Phys.\ Lett.\ }{\bf B341} (1995) 313.

\bibitem{frenchhiggs}
A.~Arhrib, G.~Moultaka,
in Proceedings of the Workshops \textit{Physics with e+ e- Linear Colliders} - Annecy,
Gran Sasso, Hamburg, February to September 1995, Part D,
ed.~P.M.~Zerwas, DESY 96-123D.

\bibitem{frenchnew} A.~Arhrib, G.~Moultaka, \texttt{hep-ph/9808317v2}.


\bibitem{LEPIIyellow} W.~Beenaker \textit{et. al.} in ``Physics at LEPII'',
   G. Altarelli, T.~S{\"o}strand, F.~Zwirner editors, \texttt{CERN 96-01},
   vol.~1, p.~79.



\end{thebibliography}
\end{document}